\documentclass[a4paper]{jpconf}
\usepackage{graphicx}
\begin{document}
\title{A study of binary constraints for seismology of $\delta$ Scuti stars}

\author{Orlagh L. Creevey}

\address{High Altitude Observatory/National Center for Atmospheric Research, 3080 Center Green, Boulder, Colorado, 80301, USA}
\address{Instituto de Astrof\'{i}sica de Canarias, 
  E-38200, Tenerife, Spain}
\ead{creevey@hao.ucar.edu}

\begin{abstract}
    Seismology of single $\delta$ Scuti stars has mainly been 
    inhibited by failing 
    to detect many of the theoretically predicted pulsation modes,
    resulting in difficulties with mode identification.
    Theoretical and observational advances have, however, 
    helped to overcome this problem,
    but the following questions then remain:
    do we know enough about the star to either 
    use the (few) identified mode(s) to probe
    the structure of the star? 
    or improve the determination of the stellar parameters?  
    It is now generally accepted that for the observed frequencies to 
    be used successfully as seismic probes for these objects, we 
    need to concentrate on stars where we can constrain the 
    number of free parameters in the problem, such as in binary systems
    or open clusters.
    The work presented here, investigates how much is gained in our 
    understanding of the star, by comparing the information we
    obtain from a single star with that of an eclipsing binary system.
    Singular Value Decomposition is the technique used to 
    explore the precision we expect in terms of stellar 
    parameters (such as mass, age and chemical composition)
    as well as how these parameter uncertainties propagate to the 
    Luminosity-Temperature (L-T) diagram.
    This work shows that the information content of the binary 
    system provides sufficient constraints on the models so that the mode 
    can be used to probe the star's structure.
    
\end{abstract}

\section{Introduction}

    $\delta$ Scuti stars are a class of pulsating stars located 
    on the Hertzsprung-Russell Diagram on or 
    around the Main Sequence and intersecting the instability strip.  
    They are 1.5 - 2.5 $M_{\odot}$ stars pulsating often in one main
    dominant oscillation mode or many lower amplitude pulsation modes.
    They have been interesting targets seismologically, because 
    the oscillation amplitudes often reach tenths of magnitudes, and 
    we 
    understand  their stellar structure relatively well, so seismology can 
    allow us to probe details of the microphysics such as energy
    transport mechanisms, convective core overshoot, as well as 
    other less well-developed theories such as rapid rotation.


    The main setbacks that $\delta$ Scuti seismology face (as well as 
    other pulsating stars) are 1) the fundamental stellar parameters are 
    not well-enough constrained to allow the few pulsation modes to 
    probe the structure, and 2) rapid rotation causes each of the 
    degrees $l$ to split into $2l+1$ $m$-modes, making mode-identification
    a difficult task \cite{gou05}. 
    Indeed these problems are not exclusive nor exhaustive.
    This has prompted authors to look towards objects where at least one 
    of these problems can be eliminated \cite{lb02, ah04, mac06, cos07}.
    Observing stars where the number of free parameters is constrained,
    such as in open clusters or multiple systems
    is a possibility for overcoming these obstacles.

    In order to use seismology to probe the interior of a star,
    the parameters of the star need to be known quite well, 
    for example, the mass should be known to 1-2\% \cite{cre07}.
    The observables from a binary system provide strict constraints
    on the parameters of the component stars.
    If the binary is an eclipsing and spectroscopic system, the 
    absolute values of the masses and radii can be extracted to 1-2\% (e.g. 
    \cite{rib99,las00}).

    The objective of this study is to find out if the uncertainties 
    in the stellar parameters can be reduced,
    so that seismology can
    be applied to those stars that exhibit one or few pulsation modes.
    We look at the particular case of a pulsating star in an eclipsing 
    binary system and compare the parameter uncertainties with those of an 
    isolated star.
    This study quantifies how well the stellar parameters can be extracted
    in various hypothetical systems.

    \section{Methods}

    The mathematical basis of this study lies in the application of
    Singular Value Decomposition (SVD) techniques to physical models.
    This technique has been used in previous studies, such as \cite{bro94,
      mm05, cre07}
    and it is being applied to many other areas of astrophysics and 
    science, because of its powerful diagnostic properties.
    Intricate details of this mathematical technique are elaborated upon in 
    the above mentioned publications as well as in \cite{pre92}. 
    Here we give the basic equations to enable an understanding of this 
    work.

    \subsection{Singular Value Decomposition}
    SVD is the decomposition of any $M\times N$  matrix {\bf D} into 
    3 components ${\bf U}$, ${\bf V^T}$ 
    and ${\bf W}$ given by ${\bf D} = {\bf U W V^T}$.
    ${\bf V^T}$ is the transpose of ${\bf V}$ which is an 
    $N \times N$ orthogonal matrix that contains the
    {\it input} basis vectors for ${\bf D}$, or the vectors
    associated with the parameter space.
    ${\bf U}$ is an $M \times N$ orthogonal matrix that contains the
    {\it output} basis vectors for ${\bf D}$, or the vectors
    associated with the observable space.
    ${\bf W}$ is a diagonal matrix that contains the
    {\it singular values} of ${\bf D}$.  

    The key element to our work is the description of the matrix {\bf D}.
    Here we define {\bf D} to be a matrix whose elements consist 
    of the partial derivatives of each
    of the observables with respect to each of the parameters of the system,
    in function of the expected measurement errors on each
    of the observables:
    \begin{equation}
      D_{ij} = \frac{\partial B_i}{\partial P_j} \epsilon_i^{-1}.
      \label{eqn:designmatrix}
      \end{equation}
    Here $B_i$ are each of the $i = 1, 2, ...M$ 
    observables of the system, with measurement or expected 
    errors $\epsilon_i$,
    and $P_j$ are each of the $j = 1,2,...,N$ free parameters of the
    system (see section \ref{sec:obspar} 
    for discussion on the observables and the 
    parameters).
    
    By writing the design matrix in function of the measurement errors,
    we provide a quantitative description of the information content 
    of each of the observables for determining the stellar parameters
    and their uncertainties.

    Supposing that we are looking for the true solution ${\bf P}_{\rm R}$ 
of the system.
    By starting from an initial close guess of the solution ${\bf P}_0$, 
    SVD can be used as an inversion technique by calculating a set 
    of parameter corrections ${\bf \delta P}$ 
    that minimizes some goodness-of-fit function:
    ${\bf \delta P = V\bar{W}^{-1}U^T \delta B}$, 
    where ${\bf \delta B}$ are the differences
    between the set of actual observations {\bf O} 
    and the calculated observables
    ${\bf B}_{\rm 0}$ given the initial parameters ${\bf P}_{\rm 0}$.
    ${\bf \bar{W}}$ is a modification of the matrix {\bf W} such that 
    the inverse of the values below a certain threshold are set to 0.
    The formal errors are comprised of the sum of all of 
    the ${\bf V}_k/w_k$, where each
    ${\bf V}_k/w_k$ describes the direction and magnitude to move
    each parameter, so that the true solution ${\bf P}_{\rm R}$ and formal 
    uncertainties can be given by
    \begin{equation}
      {\bf P}_{\rm R} = {\bf P}_{\rm 0} + 
      {\bf V\bar{W}^{-1}U^T \delta B} \left ( \pm \frac{{\bf V}_1}{w_1} 
      \pm  \frac{{\bf V}_2}{w_2} \pm ...
      \pm \frac{{\bf V}_N}{w_N} \right).
      \label{eqn:dscuti_er}
    \end{equation}
    The {\it covariance matrix} {\bf C} consequently comes 
    in a very neat and compact
    form:
    \begin{equation}
      C_{jl} = \sum_{k=1}^N \frac{V_{jk}V_{lk}}{w_{k}^2},
      \label{eqn:covariance}
    \end{equation}
    and the square 
    roots of the  diagonal elements of the covariance matrix are the 
    theoretical parameter uncertainties:
    \begin{equation}
      \sigma_j^2 =  \sum_{k=1}^N \left ( \frac{V_{jk}}{w_k} \right )^{2}.
      \label{eqn:uncertainties}
    \end{equation}
    

    \subsection{Observables, Parameters \& Models\label{sec:obspar}}

    We describe a single $\delta$ Scuti star by a set of parameters or 
    ingredients.  The ingredients for the stellar model are 
    mass $M$, age $\tau$, 
    rotational velocity $v$, initial 
    Hydrogen and heavy metals mass fraction $X$ and $Z$ where $X+Y+Z = 1$ and 
    $Y$ is initial Helium mass fraction,
    and mixing-length parameter $\alpha$ where applicable\footnote{For masses
    larger than about 1.5 $M_{\odot}$ the outer convective layer is 
    relatively thin, so 
    the observables of the star are not very
    sensitive to the value of the mixing-length parameter.}.
    The distance to the object $d$ is also included as a parameter.

    For a binary system, the additional parameters are the system properties:
    separation of components $a$, eccentricity of orbit $e$, 
    longitude of periastron $\omega$ and inclination of orbit $i$.
    Fortunately, both stellar components in a binary system
    share the parameters $\tau$, $X$ and $Z$,
    so then the individual stars differ mainly by
    $M$ and $v$.
    The parameters of the binary system of this study are given in 
    Table \ref{tab:parameters}.

    The observables are the measurable quantities of the system.
    These include things such as radius, 
    effective temperature, gravity, metallicity and parallax
    for a single star system.
    For a binary system, the observables include
    effective temperature ratio, 
    relative radii, radial velocities and orbital period.
    The Aarhus STellar Evolution Code \cite{chr82} is used to calculate the 
    stellar evolution models.  
    This code uses the stellar parameters as the input ingredients,
    and returns
    a set of global stellar properties such as radius and effective 
    temperature, 
    as well as the interior profiles of the star such as mass, 
    density and pressure.
    Oscillation frequencies for this rapidly rotating star are calculated
    using MagRot \cite{gt90, bt06}. 
    Using the global stellar properties and the distance to the star, 
    SDSS \cite{yor00} 
    magnitudes and colours in various filters are evaluated using 
    the Basel model atmospheres \cite{lej97}.
    Most of the binary observables are calculated analytically using the 
    combined stellar and binary parameters.
    In this work we refer to the non-seismic data as 
    the {\it classical} observables.
    Most of the classical observables of the binary system are given in 
    Table~\ref{tab:parameters}.

    A clear distinction should be made between the input parameters of the 
    system and output measurable quantities, the observables.
    So to discriminate between the errors in both parameters and observables,
    we shall denote the (derived) parameter uncertainties by $\sigma$,
    while $\epsilon$ is reserved for the 
    observable errors.

    Both luminosity and effective temperature can be observables, but 
    in Section \ref{sec:lt} the derived uncertainties of these properties are 
    discussed.
    This refers specifically to the calculated error boxes in the 
    luminosity-temperature (L-T) diagram, and here their uncertainties 
    will also be denoted by $\sigma$.

    \begin{table}
      \centering
      \caption{\label{tab:parameters}
	System Parameters \& Observables}
      \begin{tabular}{@{}l*{15}{l}}
	\br
	Parameter  & Value ($P_j$) & & & Observable & Value ($O_i$) 
	& $\epsilon_i$ & & \\
	\mr
	$M_A$ & 1.8 $M_{\odot}$ & & &$R_A$ & 1.95 $R_{\odot}$ & 0.02 \\
	$M_B$ & 1.7 $M_{\odot}$ & & &$R_B$ & 1.81 $R_{\odot}$ & 0.02 \\
	$\tau$ & 0.7 Gyr & & & $T_B/T_A$ & 0.97 & 0.05\\
	$X$ & 0.700 & & & $T_{\rm eff}$ & 6965 (K)  & 100 \\
	$Z$ & 0.035 & & & [M/H] & 0.31 (dex) & 0.05\\
	$v_A$ & 100.0 km s$^{-1}$  & & & $v_A \sin i$ & 99.7 km s$^{-1}$ 
	& 2.5 \\
	$v_B$ & 80.0 km s$^{-1}$ & & & $v_B \sin i$ & 59.8 km s$^{-1}$& 2.5 \\
	$d$ & 200 pc  & & & $\pi$ & 5.0 (mas) & 0.5\\
	$a$ & 0.15 AU  & & &$\Pi$ & 0.031 (yrs) & 0.00001\\
	$i$ & 85.6 $^{\circ}$ & & & $i$  & 85.6 $^{\circ}$ & 0.05\\
	$e$ & 0.0  & & & $M_A \sin^3 i $ & 1.78  $M_{\odot}$& 0.06\\
	$\omega$ & 0.0  & & & $M_B \sin^3 i$ & 1.69  $M_{\odot}$& 0.05\\
	\br
      \end{tabular}
    \end{table}

    \section{Results}

    The theoretical uncertainties in each of the 
    parameters of $M_A$, $\tau$, $X$ and $Z$ are calculated as a function 
    of error in radius $\epsilon_{R_{\star}}$ 
    and as a function of error in colour 
    $\epsilon_{(i-z)}$,
    using Equation \ref{eqn:uncertainties} coupled with the 
    observable errors given 
    in Table \ref{tab:parameters}.
    Consequently the theoretical uncertainties in luminosity and effective 
    temperature are calculated using Equation \ref{eqn:dscuti_er}, 
    for three different 
    values of $\epsilon_{R_{\star}}$ and three different values 
    of effective temperature error $\epsilon_{T_{\rm eff}}$.
    The results are shown in Figures \ref{fig1} and \ref{fig2} below.

    \subsection{Parameter Uncertainties}

\begin{figure}
      \begin{center}
	\includegraphics[width = 0.45\textwidth, height = 0.75\textheight]
			{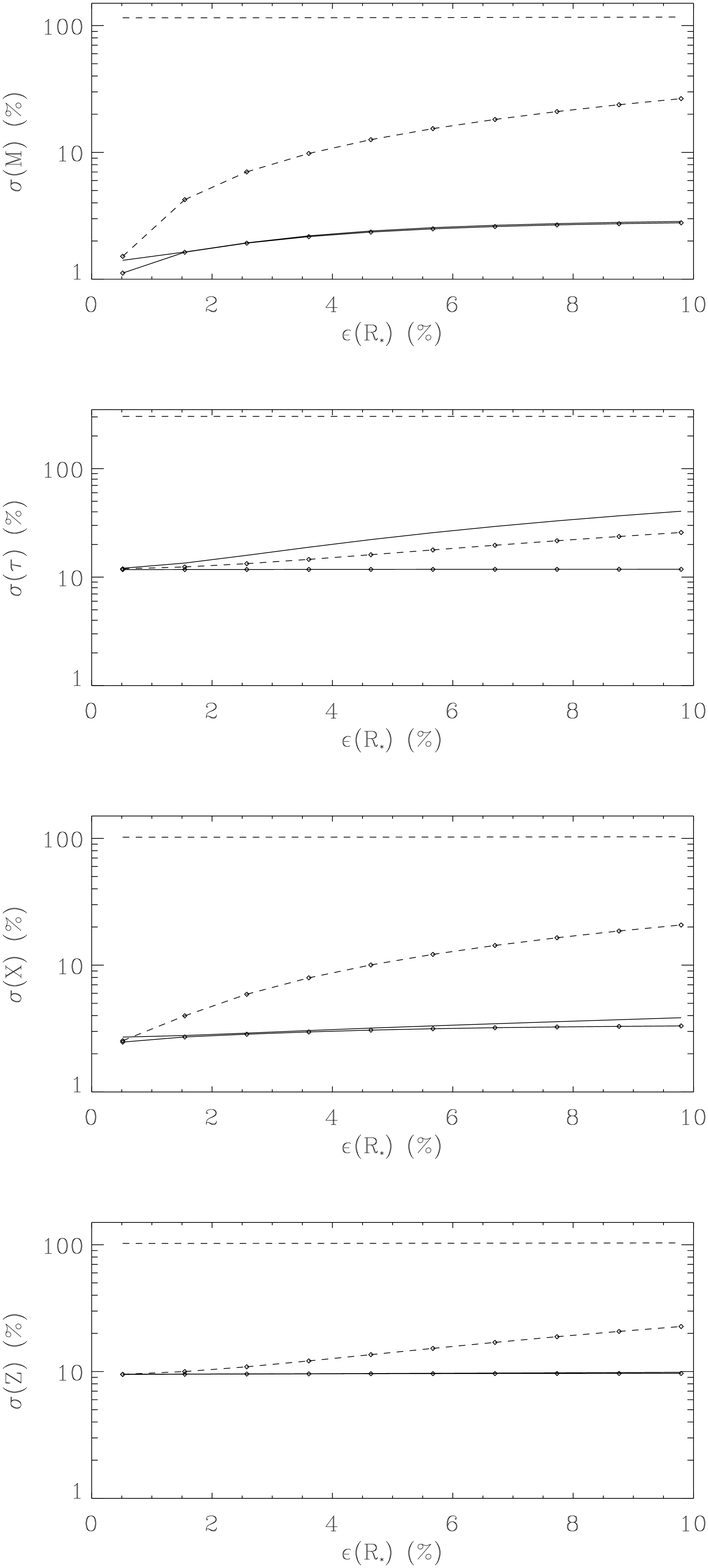}
	\includegraphics[width = 0.45\textwidth, height = 0.75\textheight]
			{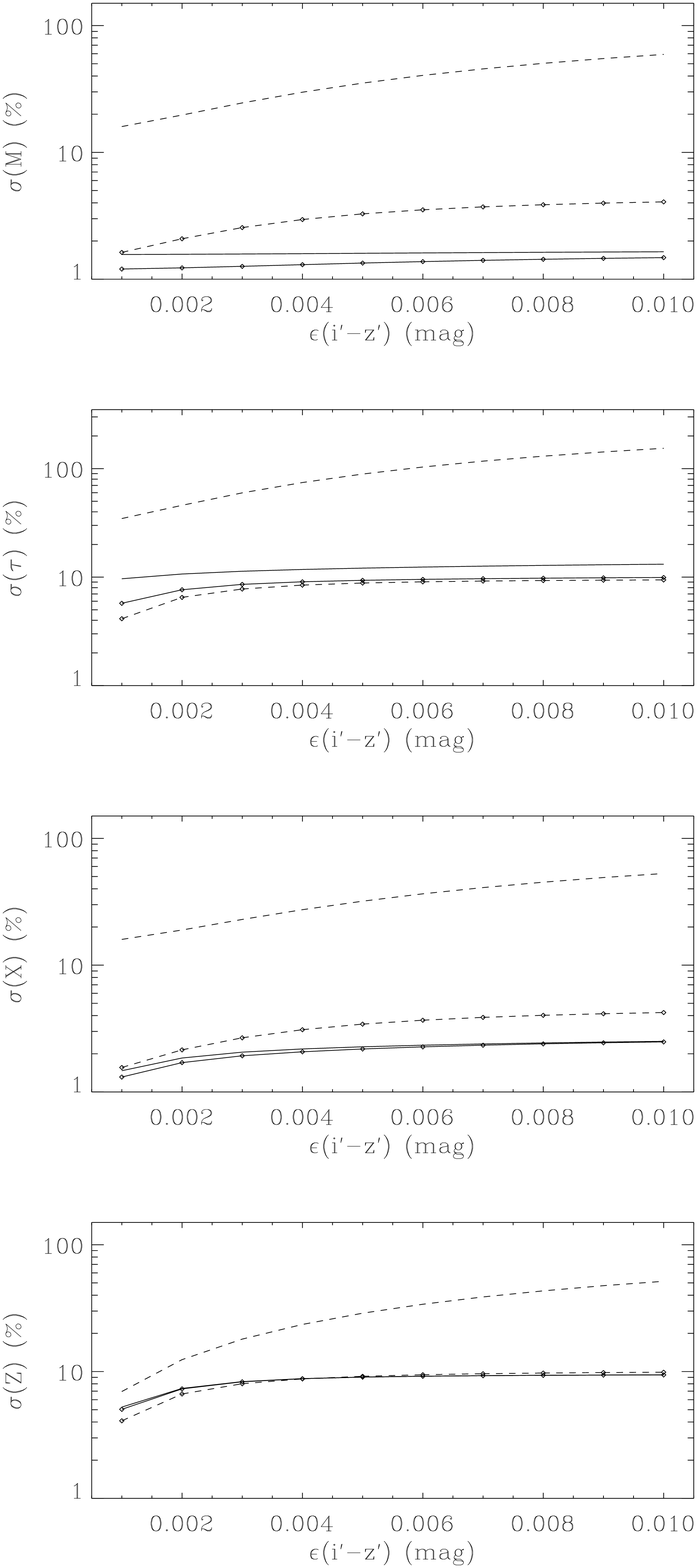}
      \end{center}
      \caption{\label{fig1} Theoretical uncertainties ($\sigma$) in mass $M$,
	age $\tau$, initial hydrogen $X$ and metal $Z$ content as a function
	of observable error.
	The left panel shows the uncertainties as a function of observable 
	radius error, here no photometric information has been included.
	The right panel shows the uncertainties as a function of observable
	error in photometric colour.
	The dashed and continuous lines show the results for the single 
	star and the binary system respectively,
	those with diamonds show the results when an identified mode
	is included in the set of observables.}
    \end{figure}

    Figure \ref{fig1} shows the theoretical uncertainties ($\sigma$) 
    in $M_A$, $\tau$, $X$ and $Z$ 
    of a pulsating star 
    as a function of error in the radius ({\it left panels}) and 
    as a function of error in the photometric colours ({\it right panels}).
    Note that the left panels do {\it not} include photometric information,
    i.e. no colours
    nor magnitudes.
    The dashed lines show the results for a single star (the observables
    are radius, effective temperature, gravity, metallicity, ...)
    while the solid lines show the results for a component of a binary 
    system (observables are those of the single star and the binary
    observables).
    The lines with the diamonds include one identified mode as well as the 
    classical observables.
    The results for the other stellar parameters are not shown, because  
    the four aforementioned parameters and $v_A$ are responsible for
    determining
    the model structure of the pulsating star.
    $\sigma (v_A)$ is  
    is usually independent of 
    $\epsilon_{R_{\star}}$ and $\epsilon_{(i-z)}$; it is determined mainly
    by the observables $v \sin i$ and $i$ from a combination of 
    spectroscopy and the 
    photometric light curve.

    For the single star without an identified mode ({\it dashed lines no 
    diamonds}), the parameter uncertainties remain at a large
    constant value as a function of radius ({\it left panels}) but do decrease
    slightly with improved photometric data ({\it right panels}).
    Only when seismic data are included for the single star system 
    ({\it dashed lines, diamonds}) {\it and} the observable
    errors are small,  the parameters are constrained to a usable amount.

    By comparing the solid lines with and without diamonds in Figure \ref{fig1}, 
    it can be seen that  
    the addition of one identified mode
    makes almost no difference to the parameter uncertainties
    for the binary system.
    This implies that there is enough information provided by the 
    binary constraints to sufficiently determine the stellar parameters.
    In this sense, the identified mode is {\it redundant} information, 
    and thus can be used maybe to test the interior of the star.

    Including photometric information ({\it right panels}) provides 
    an interesting result:
    the information provided 
    by the single star system can supersede that of the binary system for $\tau$ and $Z$ .
    This is because
    the colours are uncontaminated by a component star.
    This only happens at very small measurement errors, and only 
    when an identified mode is included for the single star.

    \subsection{Luminosity-Temperature Error Box \label{sec:lt}}

\begin{figure}
      \begin{center}
	\includegraphics[width = 1.\textwidth, height = 0.3\textheight]
			{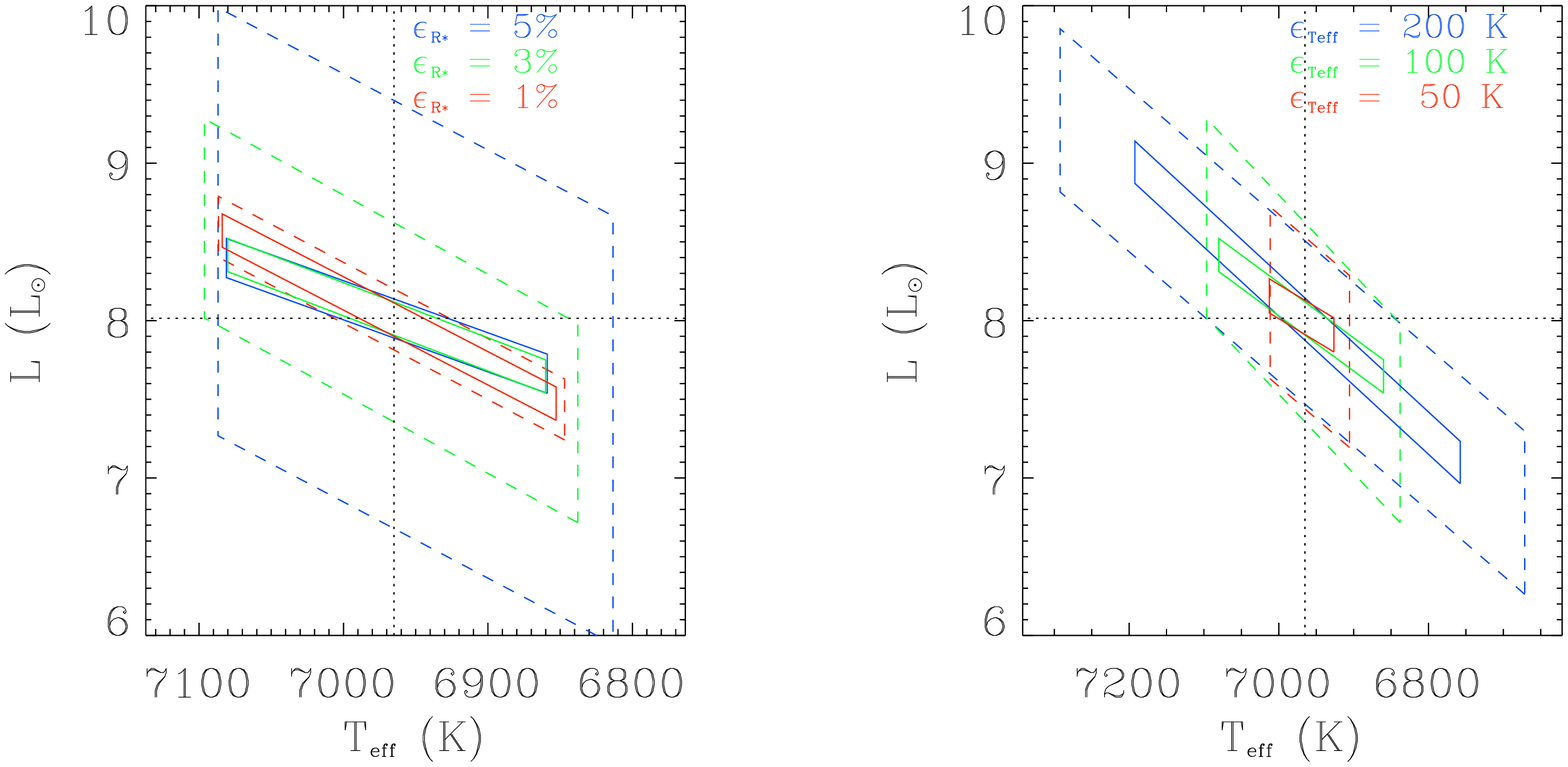}
      \end{center}
      \caption{\label{fig2} The theoretical error boxes for luminosity and 
	effective temperature.  The dashed lines represent the results for the 
	single star while the continous lines represent the results for the 
	binary system.
	The left panel shows the results while reducing the error in radius,
	and the right panel shows the results while reducing the error 
	in effective temperature.}
    \end{figure}

    The correlation matrices come in a compact form when using SVD.
    This then allows a calculation of the theoretical uncertainties
    in both effective temperature and luminosity (L-T) 
    (Equation \ref{eqn:dscuti_er}).
    Figure \ref{fig2} shows the theoretical error box in effective 
    temperature and luminosity for a single star system ({\it dashed lines})
    and a binary system ({\it solid lines}).  
    The observables do not include photometric information,
    and for the single star an identified mode is included\footnote{Figure 
      \ref{fig1} shows that the parameters are 
    not constrained for the single star if the identified mode 
    is not included.},
    while 
    for the binary system no seismic data is included.

    \subsubsection{Single Star}

    Observe how the error box reduces significantly while reducing the error 
    in the radius observable ({\it left panel}).
    The uncertainty in $T_{\rm eff}$ also reduces slightly.
    The right panel also shows that by reducing 
    the errors in the observable $T_{\rm eff}$, an expected
    corresponding reduction in the uncertainties in $T_{\rm eff}$ is noted.
    The $\epsilon_{T_{\rm eff}}$ of 200, 100, and 50 K, produces a  
    $\sigma (T_{\rm eff})$ of 250, 110, and 50 K.
    The fact that these uncertainties are reproduced also gives
    confidence in this method.
    $\sigma(L_{\star})$ changes slightly as a function of 
    $\epsilon_{T_{\rm eff}}$, its value is determined mostly by 
    the error in the radius observable (2\%). 
    Looking back to the left panel, we see that 
    interpolating between 1\% and 3\% $\epsilon_{R_{\star}}$
    produces a $\sigma(L_{\star})$ = 0.5 
    L$_{\odot}$ for $\epsilon_{R_{\star}}$ = 2\%.
    This is the value that is shown in the right panel.
    
    \subsubsection{Binary System}

    For the binary system ({\it solid lines}), {\it no} identified mode is
    included.
    The error box for the binary system does not reduce while 
    reducing the errors
    in the radius, 
    because of the small 
    uncertainties in these parameters.
    However, the error box does reduce when the error in effective temperature 
    is reduced, reproducing accurately the input $\epsilon_{T_{\rm eff}}$
    of $\sigma(T_{\rm eff})$ = 200, 100, and 50 K.
    $\sigma (L_{\star})$ does not decrease in either panel,
    because the mass is well-determined for the binary system 
    and provides this narrow constraint on $L_{\star}$.

    In all cases, note that 
    the constraints provided by the binary system without an identified 
    mode are more 
    effective than those from the single star when an identified mode is 
    included.

    \section{Conclusions}

    This study investigated whether the uncertainties in the stellar 
    parameters of a pulsating component in an eclipsing binary 
    system were sufficient so that an observed pulsation mode could 
    be used to test the physics of the stellar interiors.
    Additionally we studied the information content of a pulsating 
    star in a single star system 
    to quantify how much is gained in terms of precision in parameters
    and size of the L-T error box by observing the 
    star in a detached eclipsing binary system.
    The conclusions are summarized as follows:

    \begin{itemize}
    \item A single star system without an identified mode remains 
      poorly understood
      when observables such as the radius or colours are poorly measured.
      The parameter uncertainties are too large to correctly place
      the star in the L-T diagram.
    \item A binary system {\it without} seismic information
      provides better constraints than the single star system 
      when an oscillation 
      mode has been identified.
      \item Reducing the size of some observable errors has little or no impact
	on the parameter determinations for the binary system, 
	because these parameters are already well constrained.
    \item The tight constraints provided by the binary system for the stellar
      parameters reduces the size of the error box in the L-T 
      diagram significantly. 
      
      \item By carefully constraining the parameters of the star, just as 
	an eclipsing binary system allows us to do, 
	an accurate estimate of the stellar model under study can be obtained.
	This allows the redundant observables (like an oscillation mode) 
	to be used exclusively to test the physics of the interior
	of a star.
    \end{itemize}

\end{document}